\documentclass[graybox]{svmult}

\usepackage{mathptmx}       
\usepackage{helvet}         
\usepackage{courier}        
\usepackage{type1cm}        
\usepackage{makeidx}         
\usepackage{graphicx}        
\usepackage{multicol}        
\usepackage[bottom]{footmisc}

\makeindex             

\newcommand\la{\mathrel{\hbox{\rlap{\hbox{\lower4pt\hbox{$\sim$}}}\hbox{$<$}}}}
\newcommand\ga{\mathrel{\hbox{\rlap{\hbox{\lower4pt\hbox{$\sim$}}}\hbox{$>$}}}}

\newcommand\FeII{[Fe\,{\sc ii}]}
\newcommand\chandra{{\sl Chandra}}
\def\kms{km~s$^{-1}$}

\begin{document}

\title*{Pulsar Wind Nebulae}
\author{Patrick Slane}
\institute{Patrick Slane \at Harvard-Smithsonian Center for Astrophysics, \email{slane@cfa.harvard.edu}}
%
%
\maketitle

\abstract*{
The extended nebulae formed as pulsar winds expand into their
surroundings provide information about the composition of the winds,
the injection history from the host pulsar, and the material into
which the nebulae are expanding. Observations from across the
electromagnetic spectrum provide constraints on the evolution of
the nebulae, the density and composition of the surrounding ejecta,
the geometry of the central engines, and the long-term fate of the
energetic particles produced in these systems. Such observations
reveal the presence of jets and wind termination shocks, time-varying
compact emission structures, shocked supernova ejecta, and newly
formed dust. Here I provide a broad overview of the structure of
pulsar wind nebulae, with specific examples from observations
extending from the radio band to very high energy gamma-rays that
demonstrate our ability to constrain the history and ultimate fate
of the energy released in the spin-down of young pulsars.
}

\abstract{
The extended nebulae formed as pulsar winds expand into their
surroundings provide information about the composition of the winds,
the injection history from the host pulsar, and the material into
which the nebulae are expanding. Observations from across the
electromagnetic spectrum provide constraints on the evolution of
the nebulae, the density and composition of the surrounding ejecta,
the geometry of the central engines, and the long-term fate of the
energetic particles produced in these systems. Such observations
reveal the presence of jets and wind termination shocks, time-varying
compact emission structures, shocked supernova ejecta, and newly
formed dust. Here I provide a broad overview of the structure of
pulsar wind nebulae, with specific examples from observations
extending from the radio band to very high energy gamma-rays that
demonstrate our ability to constrain the history and ultimate fate
of the energy released in the spin-down of young pulsars.
}

\section{Introduction}
\label{sec:1}

The explosion of a supernova triggered by the collapse of a massive
star produces several solar masses of stellar ejecta expanding at
$\sim 10^4 {\rm\ km\ s}^{-1}$ into surrounding circumstellar (CSM)
and interstellar (ISM) material. The resulting forward shock
compresses and heats the ambient gas.  As the shock sweeps up
material, the deceleration drives a reverse shock (RS) \index{reverse
shock} back into the cold ejecta, heating the metal-enhanced gas
to X-ray-emitting temperatures.  In many cases, though the actual
fraction remains a currently-unsolved question, what remains of the
collapsed core is a rapidly-spinning, highly magnetic neutron star
that generates an energetic wind of particles and magnetic field
confined by the surrounding ejecta.  [All current evidence indicates
that pulsar winds are composed of electrons and positrons, with
little or no ion component. Here, and throughout, the term ``particles''
is used interchangeably for electrons/positrons.] The evolution of
this pulsar wind nebula (PWN) \index{pulsar wind nebula} is determined
by the properties of the central pulsar, its host supernova remnant
(SNR), and the structure of the surrounding CSM/ISM.

In discussing the structure and evolution of PWNe, it is important
to distinguish two important points at the outset.  First, while
PWNe have, in the past, sometimes been referred to as SNRs (most
often as a ``center-filled'' variety), they are, in fact, {\it not}
SNRs. As discussed below, PWNe are created entirely by a confined
magnetic wind produced by an energetic pulsar. At early times, the
confining material is supernova ejecta, but at later times it can
simply be the ISM. Despite being the result of a supernova explosion
(as is a neutron star), we reserve the term SNR for the structure
produced by the expanding supernova ejecta and its interaction with
the surrounding CSM/ISM (and, indeed, an entire population of SNRs
have no association with PWNe whatsoever; see Chapter ``Type Ia
supernovae''). Second, when describing the evolutionary phase of a
PWN (or a composite SNR -- an SNR that contains a PWN), it is not
necessarily the true age of the system that describes its structure.
Rather, it is the {\it dynamical} age, which accounts for the fact
that identical pulsars expanding into very different density
distributions, for example, will evolve differently.

The outline of this paper is as follows. In Section 2 we review the
basic properties of pulsars themselves, including a description of
pulsar magnetospheres and the subsequent pulsar winds that form
PWNe. Section 3 discusses the emission from PWNe and provides
examples of the constraints that multiwavelength observations place
on the determination of the system evolution.  In Section 4 we
investigate the different stages of evolution for a PWN, starting
with its initial expansion inside an SNR and ending with the bow shock
stage after the PWN escapes into the ISM.  Section 5 presents a
brief summary. Crucially, in the spirit of this Handbook, this paper
is not intended as a literature review. A small set of examples
have been selected to illustrate particular properties,
and a subset of the recent theoretical literature has been summarized
to provide the framework for our basic understanding of these
systems. The reader is referred to more thorough PWN reviews by
Gaensler \& Slane (2006), Bucciantini (2011), and Kargaltsev et al.
(2015), and to the many references and subsequent citations in those
works, for a more detailed treatment.

\section{Basic Properties}
\label{sec:2}

\subsection{Pulsars} \index{pulsars}
\label{sec:2.1}

The discovery and basic theory of pulsars has been summarized in
many places. First discovered by their radio pulsations, it was
quickly hypothesized that these objects are rapidly-rotating,
highly-magnetic neutron stars (NSs). Observations show that the
spin period $P$ of a given pulsar increases with time, indicating
a gradual decrease in rotational kinetic energy:
\begin{equation}
\dot{E} = I \Omega \dot{\Omega},
\end{equation}
where $\Omega = 2\pi/P$ and $I$ is the moment of inertia of the NS
(nominally $I = \frac{2}{5}MR^2$, where $M$ and $R$ are the mass
and radius of the star; $I \approx 10^{45}{\rm\ g\ cm}^2$ for $M =
1.4 \ M_\odot$ and $R = 10$~km). This spin-down \index{spin-down}
energy loss is understood to be the result of a magnetized particle
wind produced by the rotating magnetic star. Treated as a simple
rotating magnetic dipole, the energy loss rate is
\begin{equation}
\dot{E} = -\frac{B_pR^6\Omega^4}{6c^3}\sin^2\chi,
\end{equation}
where $B_p$ is the magnetic dipole strength at the pole and $\chi$
is the angle between the magnetic field and the pulsar rotation
axis.  Typical values for $P$ range from $\sim 0.03 - 3$s, with
period derivatives of $10^{-17} - 10^{-13} {\rm\ s\ s}^{-1}$ (though
values outside these ranges are also observed, particularly for
so-called magnetars and millisecond pulsars).  This leads to inferred
magnetic field strengths of order $10^{11} - 10^{13}$ G.


As the pulsar rotates, a charge-filled magnetosphere \index{magnetosphere}
is created, with particle acceleration occurring in charge-separated
gaps in regions near the polar cap or in the outer magnetosphere,
which extends to the so-called light cylinder \index{light cylinder}
where $R_{\rm LC} = c/\Omega$.  The maximum potential generated by
the rotating pulsar field under the assumption of co-alignment of
the magnetic and spin axes is 
\begin{equation} 
\Phi = \left(\frac{\dot{E}}{c}\right) \approx 6 \times 10^{13}
\left( \frac{\dot{E}}{10^{38}{\rm\ erg\ s}^{-1}}\right)^{1/2} {\rm\
V}.
\end{equation} 
The minimum particle current required to sustain the charge density
in the magnetosphere is
\begin{equation}
\dot{N}_{GJ} = \frac{c \Phi}{e} \approx 4 \times 10^{33}  \left(
\frac{\dot{E}}{10^{38}{\rm\ erg\ s}^{-1}}\right)^{1/2} {\rm\ s}^{-1},
\end{equation} 
where $e$ is the electron charge (Goldreich \& Julian 1969). As the
particles comprising this current are accelerated, they produce
curvature radiation that initiates an electron-positron pair cascade.
Based on observations of PWNe, values approaching $\dot{N} =
10^{40}{\rm\ s}^{-1}$ are required to explain the radio synchrotron
emission.  The implied multiplicity (i.e., the number of pairs
created per primary particle) of $\sim 10^5 - 10^7$ appears difficult
to obtain from pair production in the acceleration regions within
pulsar magnetospheres (Timokhin \& Harding 2015), suggesting that
a relic population of low energy electrons created by some other
mechanism early in the formation of the PWN may be required (e.g.,
Atoyan \& Aharonian 1996).

\subsection{Pulsar Wind Nebulae} \index{pulsar wind nebula}
\label{sec:2.2}

\begin{figure}[t]
\includegraphics[width=4.65in]{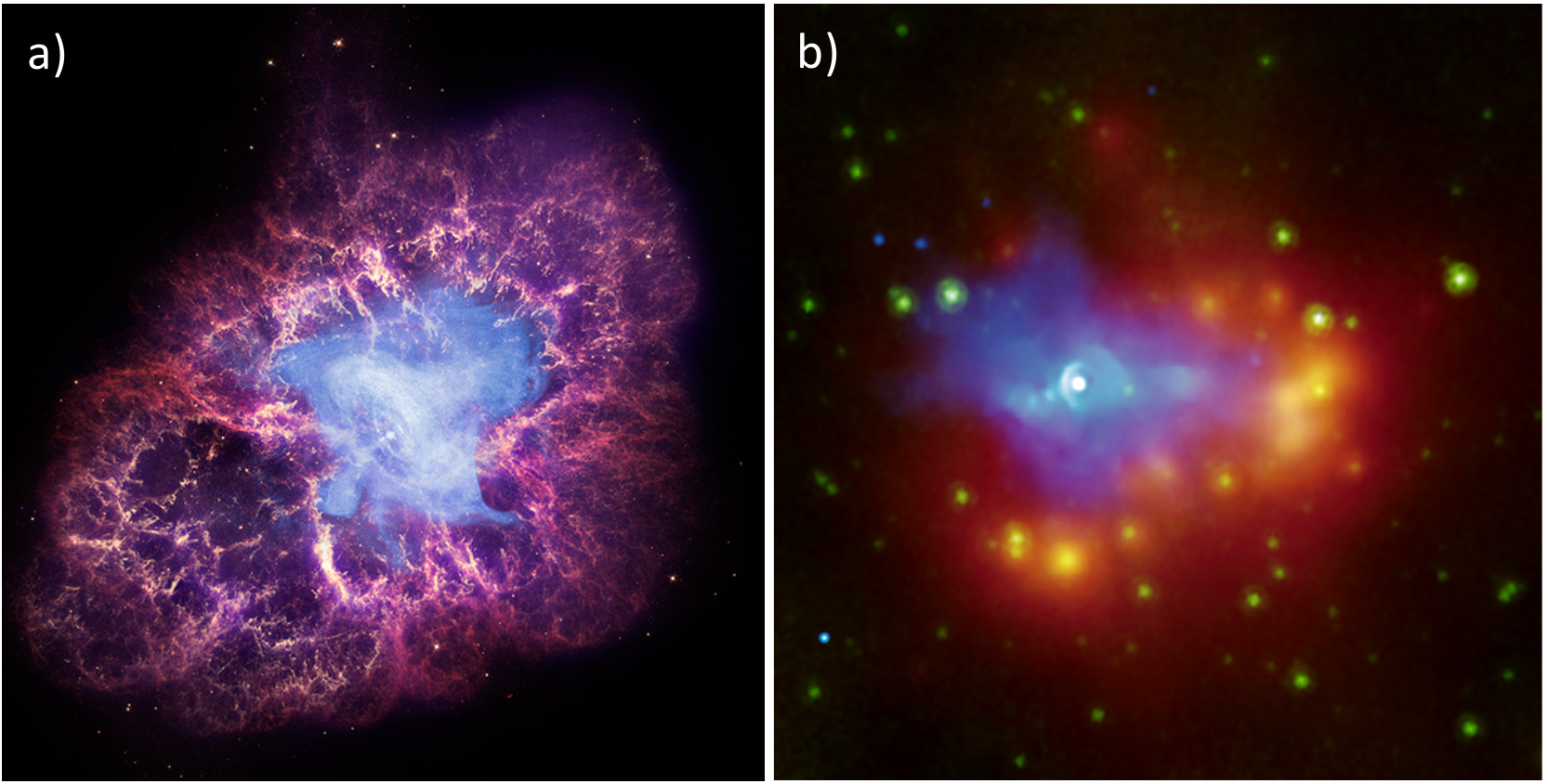}
\caption{
a) Composite image of Crab Nebula with X-ray emission (blue) from \chandra,
optical emission (red and yellow) from {\sl HST}, and IR emission (purple)
from {\sl Spitzer}. b) Composite image of G54.1+0.3 (Temim et al. 2010) 
with X-ray emisson (blue) from \chandra, and IR emission from {\sl Spitzer} 
(red-yellow, $24 \mu$m; green, $8 \mu$m). [Images courtesy NASA/CXO.]
}
\label{fig:figure1}       
\end{figure}

For pulsars with a magnetic axis that is inclined relative to the
rotation axis, the result of the above is a striped wind, with an
alternating poloidal magnetic field component separated by a current
sheet (Bogovalov 1999).  The magnetization \index{magnetization}
of the wind, $\sigma$, is defined as the ratio between the the
Poynting flux and the particle energy flux: 
\begin{equation}
\sigma = \frac{B^2}{4 \pi m n \gamma_0 c^2}, 
\label{eq_sigma}
\end{equation} 
where $B$, $n$, and $\gamma_0$ are the  magnetic
field, number density of particles of mass $m$, and bulk Lorentz
factor in the wind, respectively.  The energy density of the wind
is expected to be dominated by the Poynting flux as it leaves the
magnetosphere, with $\sigma \sim 10^4$. Ultimately, the wind is
confined by ambient material (slow-moving ejecta in the host SNR
at early times; the ISM once the pulsar has exited the SNR), forming
an expanding magnetic bubble of relativistic particles - the PWN.
As the fast wind entering the nebula decelerates to meet the boundary
condition imposed by the much slower expansion of the PWN, a wind
termination shock (TS) \index{termination shock} is formed at a
radius $R_{\rm TS}$ where the ram pressure of the wind is balanced
by the pressure within the nebula: 
\begin{equation} 
R_{\rm TS} = \sqrt{\dot{E}/(4 \pi \omega c P_{\rm PWN}}),
\end{equation} 
where $\omega$ is the equivalent filling factor for an isotropic
wind and $P_{\rm PWN}$ is the total pressure in the nebula.  The
geometry of the pulsar system results in an axisymmetric wind
(Lyubarsky 2002), forming a torus-like structure in the equatorial
plane, along with collimated jets along the rotation axis. The
higher magnetization at low latitudes confines the expansion here
to a higher degree, resulting in an elongated shape along the pulsar
spin axis for the large-scale nebula (Begelman \& Li 1992, van der
Swaluw 2003).  This structure is evident in Figure~\ref{fig:figure1}
(left), where X-ray and optical observations of the Crab Nebula
\index{Crab Nebula} clearly reveal the jet/torus structure surrounded
by the elongated wind nebula bounded by filaments of swept-up ejecta.
The innermost ring corresponds to the TS, and its radius is
well-described by Eqn. 6.  MHD models of the jet/torus structure
in pulsar winds reproduce many of the observed details of these
systems (see Bucciantini 2011 for a review).

As discussed in Section 3, the relativistic particles in the PWN
produce synchrotron radiation extending from the radio to the X-ray
band, and upscatter ambient low-energy photons (from the cosmic
microwave background, the stellar radiation field, and emission
from ambient dust) producing inverse-Compton (IC) emission in the
$\gamma$-ray band. Curiously, models of the dynamical structure and
emission properties of the Crab Nebula require $\sigma \sim 10^{-3}$
just upstream of the termination shock (Kennel \& Coroniti 1984).
Thus, somewhere between the pulsar magnetosphere and the termination
shock, the wind converts from being Poynting-dominated to being
particle-dominated. Magnetic reconnection in the current sheet has
been suggested as a mechanism for dissipating the magnetic field,
transferring its energy into that of the particles (e.g., Lyubarsky
2003).  Recent particle-in-cell simulations of relativistic shocks
show that shock compression of the wind flow can drive regions of
opposing magnetic fields together, causing the reconnection (Sironi
\& Spitkovsky 2011). As discussed in Section 3, this process can
result in a broad particle spectrum, with a power-law-like shape
$dN/dE \propto E^{-p}$ with $p \sim 1.5$. High energy particles in
the equatorial regions can diffuse upstream of the shock, generating
turbulence that supports acceleration of subsequent particles to
high energies through a Fermi-like process, potentially creating a
steeper high-energy tail with $p \sim 2.5$.  The energy range spanned
by the flat spectral region, and the maximum energy to which the
steep spectrum extends, depend on properties of the striped wind
that change with latitude, suggesting that the integrated particle
injection spectrum may be quite complex (e.g., Slane et al. 2008).
However, the maximum Lorentz factor that appears achievable is
limited by the requirement that the diffusion length of the particles
be smaller than termination shock radius; $\gamma_{max} \sim 8.3
\times 10^6 \dot{E}_{38}^{3/4} \dot{N}_{40}^{-1/2}$. This is
insufficient to explain the observed X-ray synchrotron emission in
PWNe, suggesting that an alternative picture for acceleration of
the highest energy particles in PWNe is required (Sironi et al.
2013).

\section{Radiation from PWNe}
\label{sec:3}

\begin{figure}[t]
\sidecaption
\includegraphics[width=11.5cm]{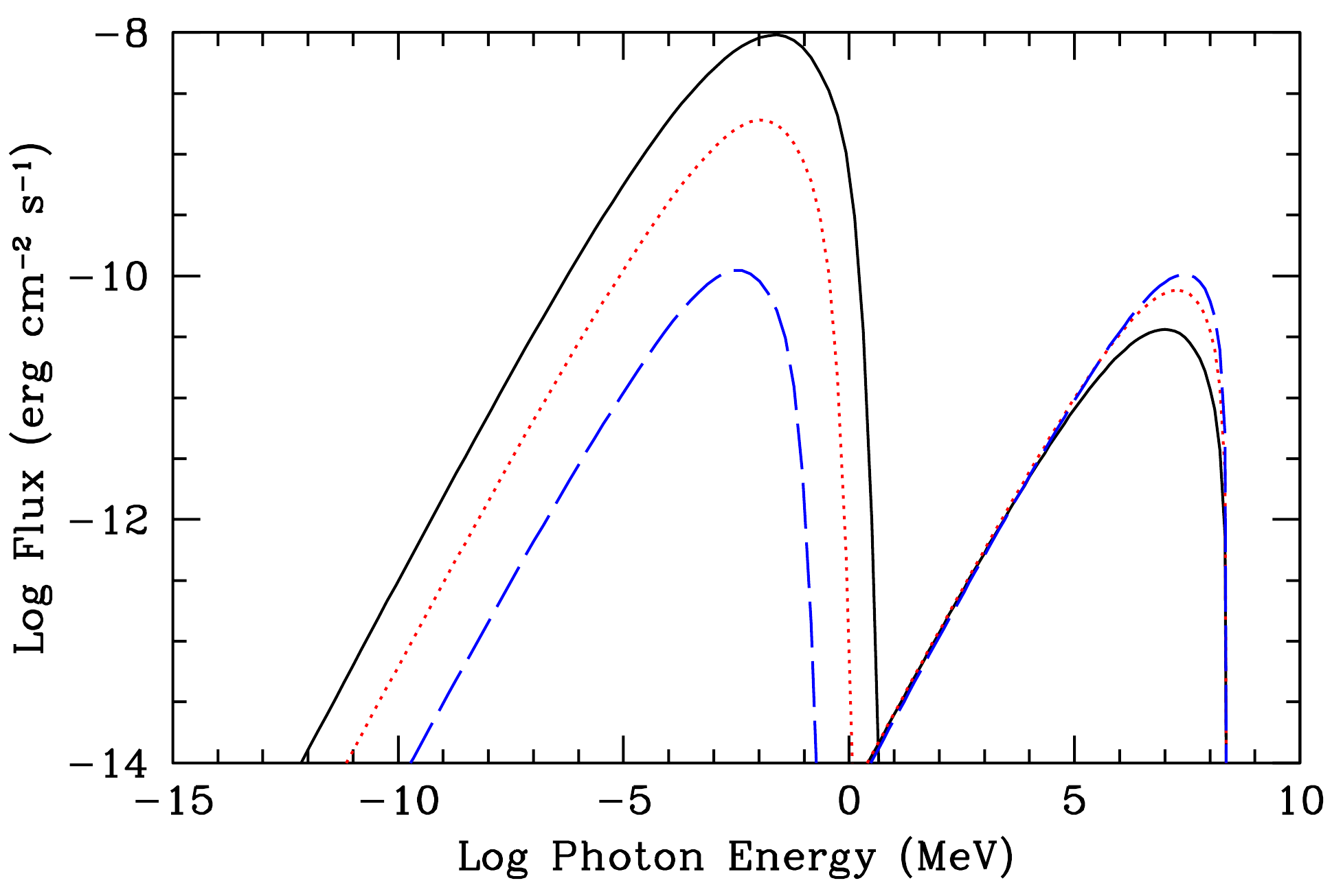}
\caption{
Synchrotron (left) and IC (right) emission (for scattering off of
the CMB) from a PWN at ages of 1000 (solid), 2000 (dotted), and
5000 (dashed) years. Here we have assumed $E_{51} = 1$, $M_{ej} =
8 M_\odot$, and $n_0 = 0.1 {\rm\ cm}^{-3}$ for the SNR evolution,
and $n = 3$, $\dot{E}_0 = 10^{40}{\rm\ erg\ s}^{-1}$, and $\tau_0
= 500$~yr for the pulsar. For the wind, we assume that 99.9\% of
the energy is in the form of electrons/positrons with a power law
spectrum with $\gamma$ = 1.6.
}
\label{fig:figure2}       
\end{figure}

The emission from PWNe can be divided into two broad categories --
that originating from the relativistic particles within the nebula
and that produced by material that has been swept up by the nebula.

\subsection{Emission from nebula} \index{PWN emission}
\label{sec:3.1}

The emission from the relativistic particles is a combination of
synchrotron radiation \index{synchrotron radiation} and IC radiation
\index{inverse-Compton radiation} associated with the upscattering
of ambient photons.  If we characterize the injected spectrum as a
power law, 
\begin{equation} 
Q(E_e,t) = Q_0(t)(E_e/E_0)^{-\gamma}
\end{equation} 
the integrated particle energy is then 
\begin{equation}
\int Q(E,t) E dE = (1 + \sigma) \dot{E}(t) 
\end{equation} 
The resulting emission spectrum is found by integrating the electron
spectrum over the emissivity function for synchrotron and IC radiation
using, respectively, the nebular magnetic field and spectral density
of the ambient photon field. As noted above, the low energy particles
in PWNe actually appear to have a flatter spectrum, leading to a
flat radio spectrum ($\alpha \sim 0.0 - 0.3$ where $S_\nu \propto
\nu^{-\alpha})$ [Note: In X-rays, it is conventional to express the
photon spectrum $dN_\gamma/dE \propto E^{-\Gamma}$, where $\Gamma
= \alpha + 1$.] The spectrum generally steepens near the mm or
optical band. For young PWNe with very high magnetic fields,
up-scattering of the high energy synchrotron spectrum can produce
$\gamma$-ray photons through so-called synchrotron self-Compton
emission. The resulting spectrum thus depends on the age, magnetic
field, and pulsar spin-down power (e.g., Torres et al. 2013).

As illustrated in Figure~\ref{fig:figure2},
the build-up of particles in the nebula results in an IC spectrum
that increases with time.  The synchrotron flux decreases with time
due to the steadily decreasing magnetic field strength associated
with the adiabatic expansion of the PWN (see Section 4). This
behavior is reversed upon arrival of the SNR RS (not
shown in Figure 2), following which the nebula is compressed and the
magnetic field strength increases dramatically, inducing an episode
of rapid synchrotron losses. Upon re-expanding, however, IC emission
again begins to increase relative to the synchrotron emission.  At
the latest phases of evolution, when the nebula is very large and
the magnetic field is low, the IC emission can provide the most
easily-detected signature. As described below, such behavior is
seen for a number of PWNe that have been identified based on their
emission at TeV energies, and for which only faint synchrotron
emission near the associated pulsars is seen in the X-ray band.

For electrons with energy $E_{e,100}$, in units of 100~TeV,
the typical energy of synchrotron photons is
\begin{equation}
E_\gamma^s \approx 2.2 E_{e,100}^2 B_{10} {\rm\ keV},
\label{eqn:E_syn}
\end{equation}
where $B_{10}$ is the magnetic field strength in units of $10 \
\mu$G. The associated synchrotron lifetime \index{synchrotron
lifetime} for the particles is
\begin{equation}
\tau_{syn} \approx 820 E_{e,100}^{-1} B_{10}^{-2} {\rm\ yr}
\end{equation}
which results in a break \index{synchrotron break} in the 
photon spectrum at
\begin{equation}
E_{\gamma,br} \approx 1.4 B_{10}^{-3} t_{\rm kyr}^{-2} {\rm\ keV}
\end{equation}
for electrons injected over a lifetime $t_{\rm kyr}$. Beyond this
energy, the photon power law spectrum steepens by $\Delta \Gamma =
0.5$. For young PWNe, with large magnetic fields, the result is a
steepening of the X-ray spectrum with radius due to synchrotron
burn-off \index{synchrotron burn-off} of the higher energy particles
on timescales shorter than their transit time to the outer portions
of the PWN.  This is readily observed in young systems such as
G21.5$-$0.9 and 3C~58 (see below), although the spectral index
actually flattens more slowly than expected unless rapid particle
diffusion is in effect (Tang \& Chevalier 2012).

For $\gamma$-rays produced by IC-scattering off of the CMB,
\begin{equation}
E_\gamma^{IC} \approx 0.32 E_{e,10}^2 {\rm\ TeV},
\label{eqn:E_IC}
\end{equation}
where $E_{e,10} = E_e/(10 {\rm\ TeV})$. Note that while the synchrotron
energy depends upon both the electron energy and the magnetic field
strength, the IC energy (from CMB scattering) depends only on the
particle energy. Modeling of both emission components for a particular
PWN thus allows determination of the magnetic field strength.

Because of the short synchrotron lifetime for the X-ray emitting 
particles, the X-ray luminosity is related to the current spin-down
power of the pulsar. From a variety of studies, $L_x \sim 10^{-3}
\dot{E}$ (e.g., Possenti et al. 2002). Although flux values for
individual pulsars may differ from this relationship by as much
as a factor of 10, determination of the X-ray luminosity can
provide a modest constraint on $\dot{E}$ for systems in which
pulsations are not directly detected.

The broadband spectrum of a PWN, along with the associated dynamical
information provided by measurements of the pulsar spin properties,
and the size of the PWN and its SNR, place very strong constraints
on its evolution and on the spectrum of the particles injected from
the pulsar. Combined with estimates of the swept-up ejecta mass,
this information can be used to probe the properties of the progenitor
star and to predict the long-term fate of the energetic particles
in the nebula. Recent multiwavelength studies of PWNe, combined
with modeling efforts of their evolution and spectra, have provided
unique insights into several of these areas.

\subsection{Emission from shocked ejecta} \index{SNR ejecta}
\label{sec:3.2}

As the PWN expands into the surrounding supernova ejecta, as described
below, it heats the ejecta. The resulting emission, often confined
to filaments, is a combination of radiation from shocked gas and
continuum emission from dust condensed from the cold ejecta in the
early adiabatic expansion of the SNR. The thermal emission depends
on the velocity of the PWN shock driven into the ejecta which, in
turn, depends on the spin-down power of the central pulsar and the
density and velocity profile of the ejecta. For slow shocks, line
emission may be observed in the IR and optical bands, such as that
observed from the Crab Nebula (see Chapter ``Supernova of 1054 and
its remnant, the Crab Nebula''), G21.5$-$0.9, and G54.1+0.3 (see
below), while for faster shocks the emission may appear in the X-ray
band, as observed in 3C~58.  This line emission can provide important
information on the ejecta composition and expansion velocity.

The dust emission is in the form of a blackbody-like spectrum
whose properties depend on the temperature, composition, and grain-size
distribution of the dust. Measurements of emission from ejecta dust
prior to interaction with the SNR RS (see below) are of
particular importance in estimating dust formation rates in supernovae
(e.g., Temim et al. 2015)

\section{PWN Evolution} \index{PWN evolution}
\label{sec:4}

\begin{figure}[t]
\sidecaption
\includegraphics[height=2.25in]{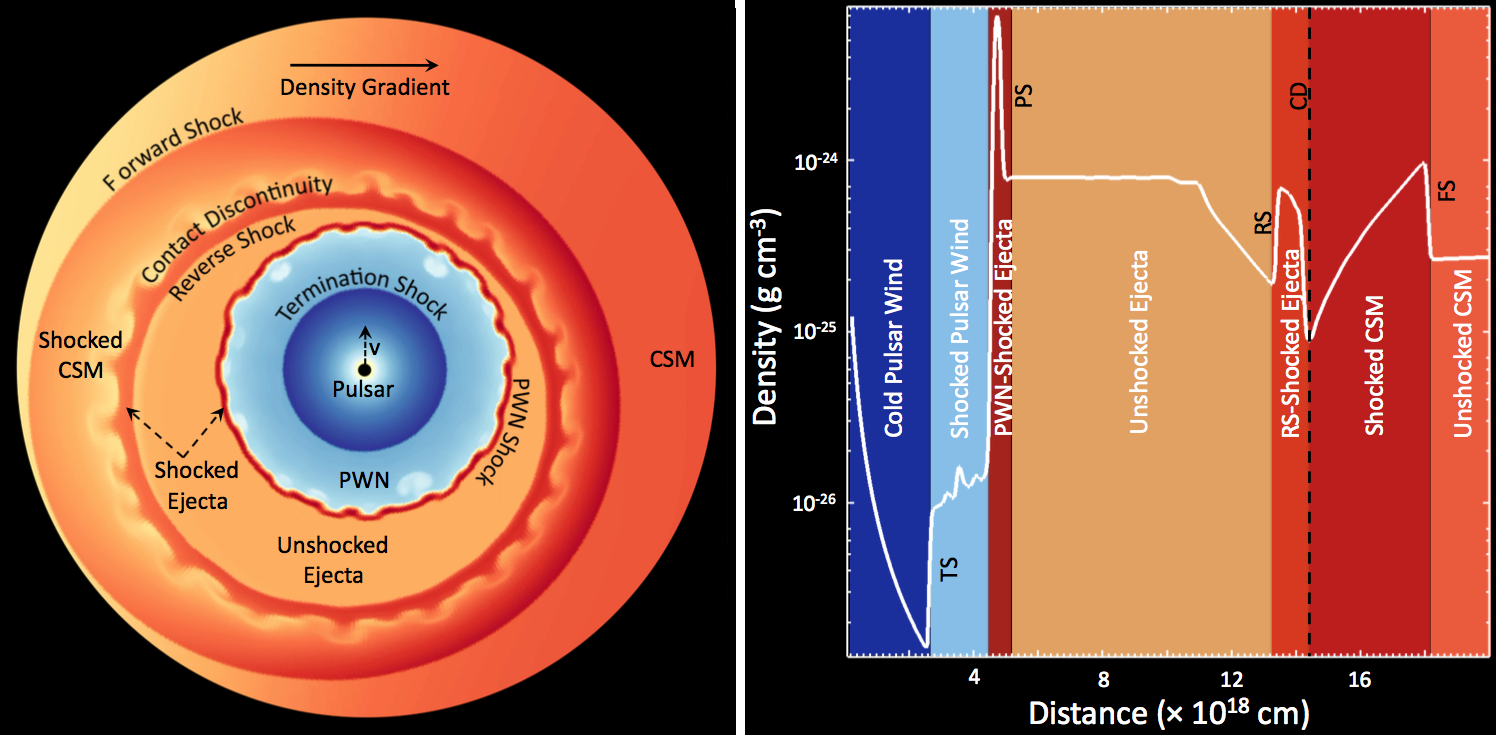}
\caption{
Left: Density image from a hydrodynamical simulation of a PWN
expanding into an SNR that is evolving into a medium with a CSM
density gradient increasing to the right. The pulsar itself is
moving upward. The reverse shock is propagating inward, approaching
the PWN preferentially from the upper right due to the combined
effects of the pulsar motion and the CSM density gradient.
Right: Density profile for a radial slice through the simulated
composite SNR. Colored regions correpond to different physical
regions identified in the SNR image.
}
\label{fig:figure3}       
\end{figure}

The evolution of a PWN within the confines of its host SNR is
determined by both the rate at which energy is injected by the
pulsar and by the density structure of the ejecta material into
which the nebula expands.  The location of the pulsar itself,
relative to the SNR center, depends upon any motion given to the
pulsar in the form of a kick velocity during the explosion, as well
as on the density distribution of the ambient medium into which the
SNR expands. At the earliest times, the SNR blast wave expands
freely at a speed of $\sim (5-10)\times10^3$~\kms, much higher than
typical pulsar velocities of $\sim 200-1500$~\kms. As a result, for
young systems the pulsar will always be located near the SNR center.

The energetic pulsar wind is injected into the SNR interior, forming
a high-pressure bubble that expands supersonically into the surrounding
ejecta, forming a shock.  The input luminosity is generally assumed
to have the form (e.g. Pacini \& Salvati 1973)
\begin{equation}
\dot{E} = \dot{E}_0 \left( 1 + \frac{t}{\tau_0}
\right)^{-\frac{(n+1)}{(n-1)}}
\label{eqn_edot_vs_t}
\end{equation}
where
\begin{equation}
\tau_0 \equiv \frac{P_0}{(n-1)\dot{P}_0}
\label{eqn_tau0}
\end{equation}
is the initial spin-down time scale of the pulsar.  Here $\dot{E_0}$
is the initial spin-down power, $P_0$ and $\dot{P}_0$ are the initial
spin period and its time derivative, and $n$ is the so-called
``braking index'' \index{braking index} of the pulsar ($n = 3$ for
magnetic dipole spin-down).  The pulsar has roughly constant energy
output until a time $\tau_0$, beyond which the output declines
fairly rapidly with time.

Figure \ref{fig:figure3} illustrates the evolution of a PWN within
its host SNR.  The left panel shows a hydrodynamical simulation of
an SNR evolving into a non-uniform medium, with a density gradient
increasing from left to right. The pulsar is moving upward. The SNR
forward shock (FS), RS and contact discontinuity (CD) separating
the shocked CSM and shocked ejecta are identified, as is the PWN
shock driven by expansion into the cold ejecta. The right panel
illustrates the radial density distribution, highlighting the PWN
TS as well as the SNR FS, CD, and RS.

\subsection{Early Expansion}
\label{sec:4.1}

The energetic pulsar wind injected into the SNR interior forms a
high-pressure bubble that drives a shock into the surrounding ejecta.
The sound speed in the relativistic fluid within
the PWN is sufficiently high ($c_s = c/\sqrt{3}$) that any pressure
variations experienced during the expansion are quickly balanced
within the bubble; at early stages, the pulsar thus remains located
at the center of the PWN, even if the pulsar itself is moving through
the inner SNR, which is often the case because pulsars can be born
with high velocities ($\sim 200 - 1500 {\rm\ km\ s}^{-1}$; Arzoumanian
et al. 2002) due to kicks imparted in the supernova explosions.
The wind is confined by the innermost slow-moving ejecta, and the
PWN expansion drives a shock into these ejecta, heating them and
producing thermal emission.  Magnetic tension in the equatorial
regions exceeds that elsewhere in the nebula, resulting an a oblate
morphology with the long axis aligned with the pulsar rotation axis
(Begelman \& Li 1992).  As illustrated in Figure~\ref{fig:figure3}
(left), the PWN/ejecta interface is susceptible to Rayleigh-Taylor
(R-T) instabilities. These structures are readily observed in the
Crab Nebula (Figure~\ref{fig:figure1}a; also see Hester 2008 as
well as Chapter ``Supernova of 1054 and its remnant, the Crab
Nebula''), where highly-structured filaments of gas and dust are
observed in the optical and infrared.  Spectral studies of these
filaments provide information on the composition, mass, and velocity
of the ejecta.  This, along with information about the associated
SNR, can place strong constraints on the progenitor system.

In the Crab Nebula, for example, the total mass of the ejecta swept
up by the PWN is $\sim 5 M_\odot$ (Fesen et al. 1997), and the
expansion velocity is $\sim 1300{\rm\ km\ s}^{-1}$ (Temim et al.
2006). The Crab is one of a small set of young PWNe for which there
is no evidence of the surrounding SNR, other than the swept-up
ejecta. Other examples include 3C~58 and perhaps G54.1$+$0.3,
although there is some evidence for radio and X-ray emission that
might be associated with an SNR shell in the latter (Bocchino et
al. 2010). The lack of bright (or any) SNR emission in these systems
is generally assumed to result from some combination of low explosion
energy, as might result from low-mass progenitors that produce
electron-capture SNe, and a very low surrounding
density, as could result from mass loss through stellar winds in
the late phase of massive star evolution.

For the Crab Nebula, the available evidence appears to be consistent
with a low-mass progenitor (Yang \& Chevalier 2015). For G54.1$+$0.3,
on the other hand, an infrared shell surrounding the X-ray PWN is
observed to encompass a collection of what appear to be O-type stars
that presumably formed in the same stellar cluster as the PWN
progenitor, indicating that this system resulted from a high mass
star (Temim et al. 2010). The IR emission appears to arise from a
combination of slow shocks driven into the surrounding ejecta and
unshocked supernova dust that is being radiatively heated by emission
from the embedded stars.

\begin{figure}[t]
\includegraphics[width=4.65in]{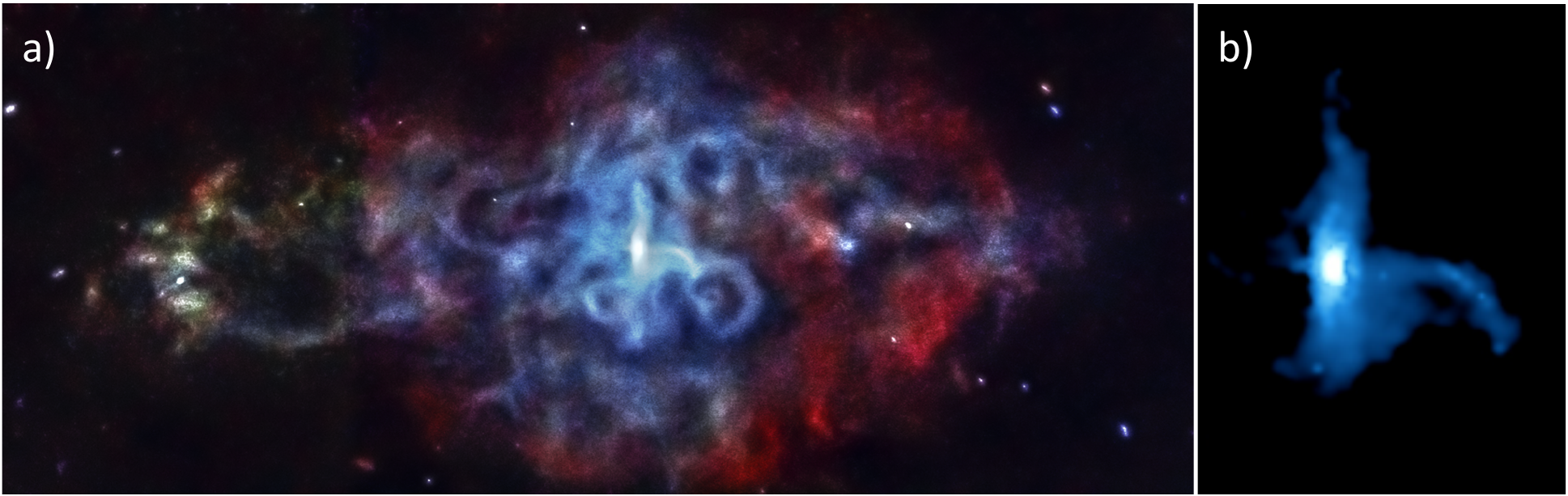}
\caption{a) \chandra\ image of 3C~58 (Slane et al. 2004). Low (high)
energy X-rays are shown in red (blue). b) Expanded view of the
central region of 3C~58 showing the toroidal structure and jet 
associated with the central pulsar. [Images courtesy NASA/CXO.]
}
\label{fig:figure4}       
\end{figure}

While the optical emission from 3C~58 \index{3C 58} shows evidence
for R-T structures, high resolution X-ray observations show a network
of filamentary structures that do not appear to be associated with
the optical filaments (Figure~\ref{fig:figure4}).  The origin of
these structures is currently not understood, although kink
instabilities \index{kink instabilities} in the termination shock
region may result in magnetic structures whose size scale is similar
to what is observed in 3C~58 (Slane et al. 2004). Thermal X-ray
emission is observed in the outer regions of the PWN, (which appear
red in Figure~\ref{fig:figure4} due to both the low energy thermal
flux and the steepening of the synchrotron spectrum with radius
associated with burn-off of high energy particles) with indications
of enhanced metals as would be expected from shocked ejecta. Mass
and abundance measurements, combined with expansion measurements,
can provide the velocity and composition distribution of the ejecta,
placing constraints on the total ejecta mass and explosion energy
of the supernova (e.g., Yang \& Chevalier 2015, Gelfand et al.
2015).

\begin{figure}[t]
\includegraphics[width=4.65in]{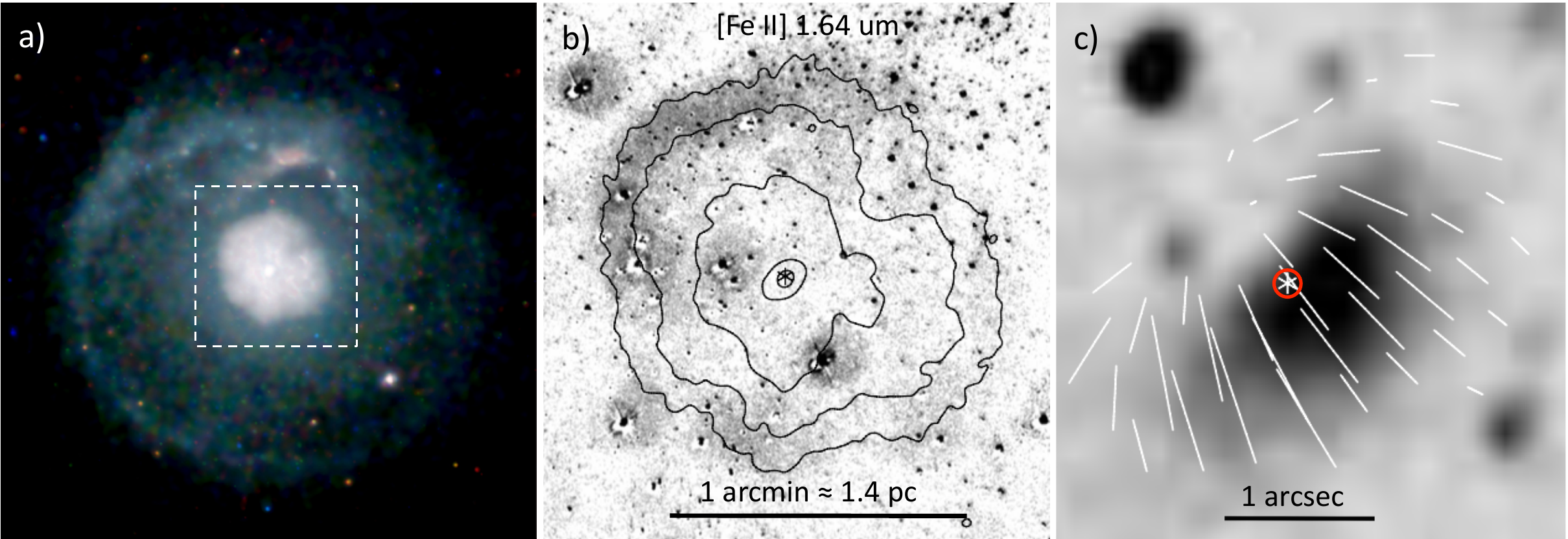}
\caption{a) \chandra\ image of G21.5$-$-.9. The pulsar is located
at the center and is surrounded by a PWN. The faint outer shell is
the SNR, and a portion of the emission between the PWN and the
outer shell is scattered flux from the PWN. b) Infrared image at
1.64 $\mu$m showing a shell of \FeII\ emission from ejecta that has
been swept up and shocked by the expanding PWN. c) Infrared
polarization image. The pulsar is within the red circle. White bars
show the direction of the electric field vectors, with the length
proportional to the polarization fraction. The inferred magentic
field, which is orthogonal to the electric vectors, is largely
toroidal. [From Zajczyk et al. 2012, A\&A, 542, A12 -
Reproduced with permission from Astronomy \& Astrophysics, 
\textcopyright ESO]
}
\label{fig:figure5}       
\end{figure}

For more typical systems, the ambient density (and/or supernova
explosion energy) is sufficiently high to form a distinct SNR shell
of swept-up CSM/ISM material, accompanied by RS-shocked ejecta, as
illustrated in Figure~\ref{fig:figure3}.  An exceptional example
is G21.5$-$0.9.  \index{G21.5$-$0.9} X-ray observations
(Figure~\ref{fig:figure5}a) show a bright central nebula that
coincides with a flat-spectrum radio nebula. The nebula is accompanied
by a faint SNR shell (Slane et al. 2000; Matheson \& Safi-Harb
2005), and radio timing measurements with the Parkes telescope
reveal the 62~ms pulsar J1833-1034 in the center of the nebula
(Camilo et al. 2006).  Ground-based IR observations (Zajczyk et al.
2012) reveal a ring of \FeII\ emission associated with ejecta that
has been swept up by the expanding PWN (Fig. 4b; contours are X-ray
emission from the dashed square region from Fig. 4a). The emission
directly around the pulsar is extended in X-rays (see innermost
contours), possibly associated with a surrounding torus as is seen
in the Crab Nebula and other PWNe.  The IR emission surrounding the
pulsar is polarized.  The electric field vectors are shown in Fig.
4c, with the length of the white bars proportional to the polarization
fraction. The magnetic field, which is perpendicular to the electric
vectors, is largely toroidal, consistent with the picture of wound-up
magnetic flux from the spinning pulsar, as described above.

\subsection{Reverse-shock Interaction} \index{reverse shock}
\label{sec:4.2}

As the SNR blast wave sweeps up increasing amounts of material, the
RS propagates back toward the SNR center. In the absence of a central
PWN, it reaches the center at a time $t_{c} \approx 7
(M_{ej}/10~M_\odot)^{5/6} E_{51}^{-1/2} n_0^{-1/3}~{\rm kyr}$, where
$E_{51}$ is the explosion energy, $M_{ej}$ is the ejecta mass, and
$n_0$ is the number density of ambient gas (Reynolds \& Chevalier
1984). When a PWN is present, however, the RS interacts with the
the nebula before it can reach the center (Figure~\ref{fig:figure3}).
The shock compresses the PWN, increasing the magnetic field strength
and resulting in enhanced synchrotron radiation that burns off the
highest energy particles. In the simplified case of SNR expansion
into a uniform medium, with a spherically-symmetric PWN, the system
evolves approximately as illustrated in Figure~\ref{fig:figure2}
(from Gelfand et al. 2009), where the Sedov solution \index{Sedov
solution} has been assumed for the SNR evolution,
\begin{equation} R_{SNR} \approx 6.2 \times
10^4 \left(\frac{E_{SN}}{n_0}\right)^{1/5} t^{2/5}, 
\end{equation}
and the PWN evolves approximately as 
\begin{equation} R_{PWN}
\approx 1.5 \dot{E}_0^{1/5} E_{SN}^{3/10} M_{ej}^{-1/2} t^{6/5}
\end{equation} 
(Chevalier 1977) prior to the RS interaction. [In
reality, the SNR expands freely at the outset, approaching the Sedov
solution as $t \rightarrow t_c.$] Here, $E_{SN}$ is the supernova
explosion energy, $n_0$ is the number density of the ambient medium,
and $M_{ej}$ is the mass of the supernova ejecta.

\begin{figure}[t]
\sidecaption
\includegraphics[width=11.5cm]{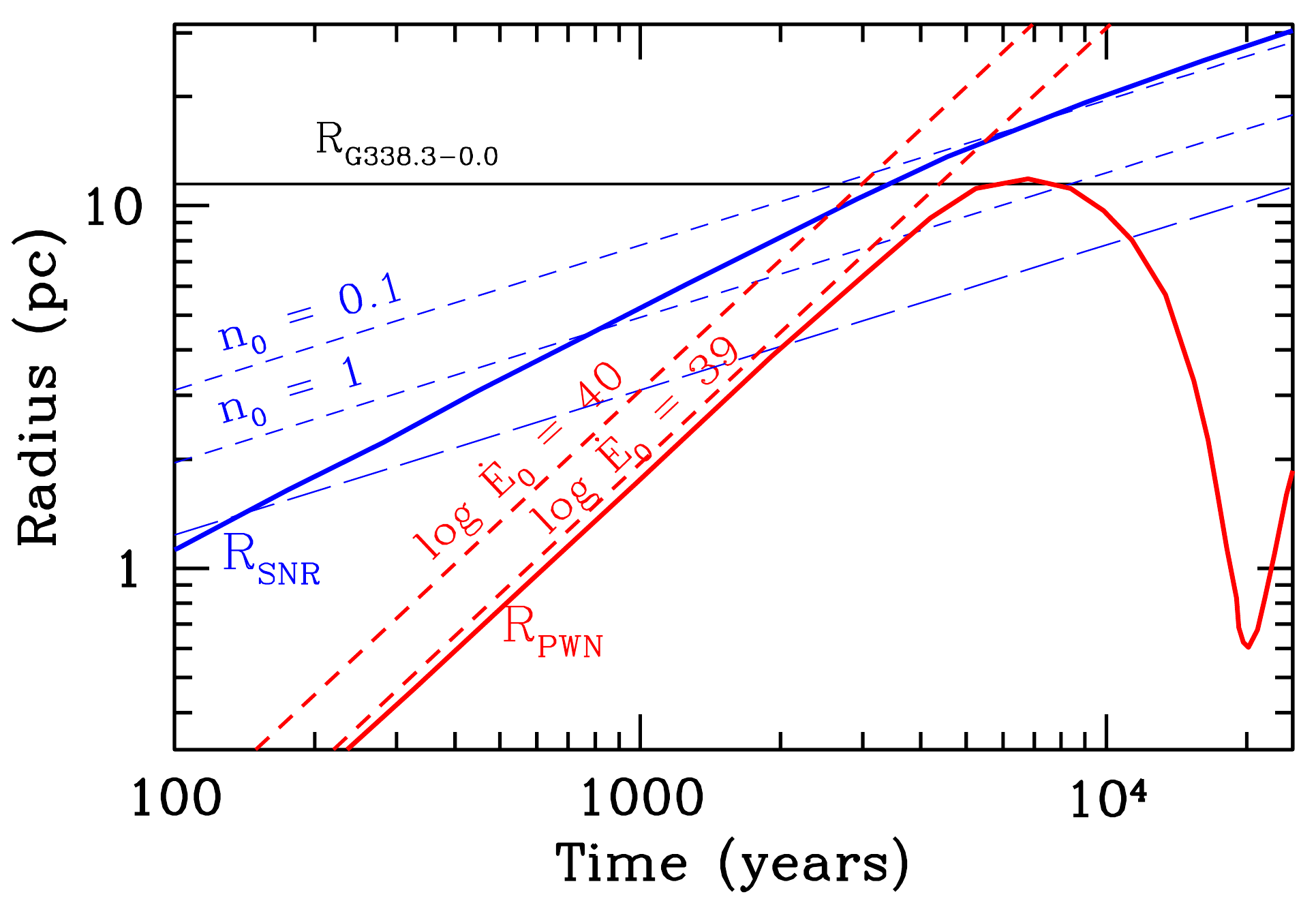}
\caption{
Time evolution of the SNR and PWN radii for a range of values for
the ambient density and initial spin-down power of the pulsar. The
solid curves correspond to models from Gelfand et al. (2009)
using
$\dot{E_0} = 10^{40} {\rm\ erg\ s}^{-1}$, $M_{ej} = 8 M_\odot$, $n_0
= 0.1 {\rm\ cm}^{-3}$, and $E_{51} = 1$. 
}
\label{fig:figure6}       
\end{figure}

If the ambient CSM/ISM is significantly nonuniform (and it
typically is, because massive stars form in turbulent regions of
dense clouds, and strongly modify the CSM through strong
and potentially-asymmetric winds), the FS expands more
(less) rapidly in regions of lower (higher) density.  This has two
significant effects.  First, it changes the morphology of the SNR
to a distorted shell for which the associated pulsar is no longer
at the center.  Second, the RS also propagates asymmetrically,
reaching the center more quickly from the direction of the higher
density medium (Blondin et al. 2001).

The return of the RS ultimately creates a collision with the PWN.
During the compression phase, \index{compression phase} the magnetic
field of the nebula increases, resulting in enhanced synchrotron
radiation and significant radiative losses from the highest energy
particles.  The PWN/RS interface is Rayleigh-Taylor (R-T) unstable,
and is subject to the formation of filamentary structure where the
dense ejecta material is mixed into the relativistic fluid.  If the
SNR has evolved in a nonuniform medium, an asymmetric RS will form,
disrupting the PWN and displacing it in the direction of lower
density (Figure~\ref{fig:figure3}).  The nebula subsequently re-forms
as the pulsar injects fresh particles into its surroundings, but a
significant relic nebula of mixed ejecta and relativistic gas will
persist.


Because the SNR RS typically reaches the central PWN on a timescale
that is relatively short compared with the SNR lifetime, all but
the youngest PWNe that we observe have undergone an RS interaction
(see Figure~\ref{fig:figure6}). This has significant impact on the
large-scale geometry of the PWN, as well as on its spectrum and
dynamical evolution.  Remnants such as G328.4+0.2 (Gelfand et al.
2007), MSH 15$-$56 (Temim et al. 2013), and G327.1$-$1.1 (Temim et
al. 2015) \index{G327.1$-$1.1} all show complex structure indicative
of RS/PWN interactions, and observations of extended sources of
very high energy (VHE) $\gamma$-rays indicate that many of these
objects correspond to PWNe that have evolved beyond the RS-crushing
stage.

\begin{figure}[t]
\includegraphics[width=4.65in]{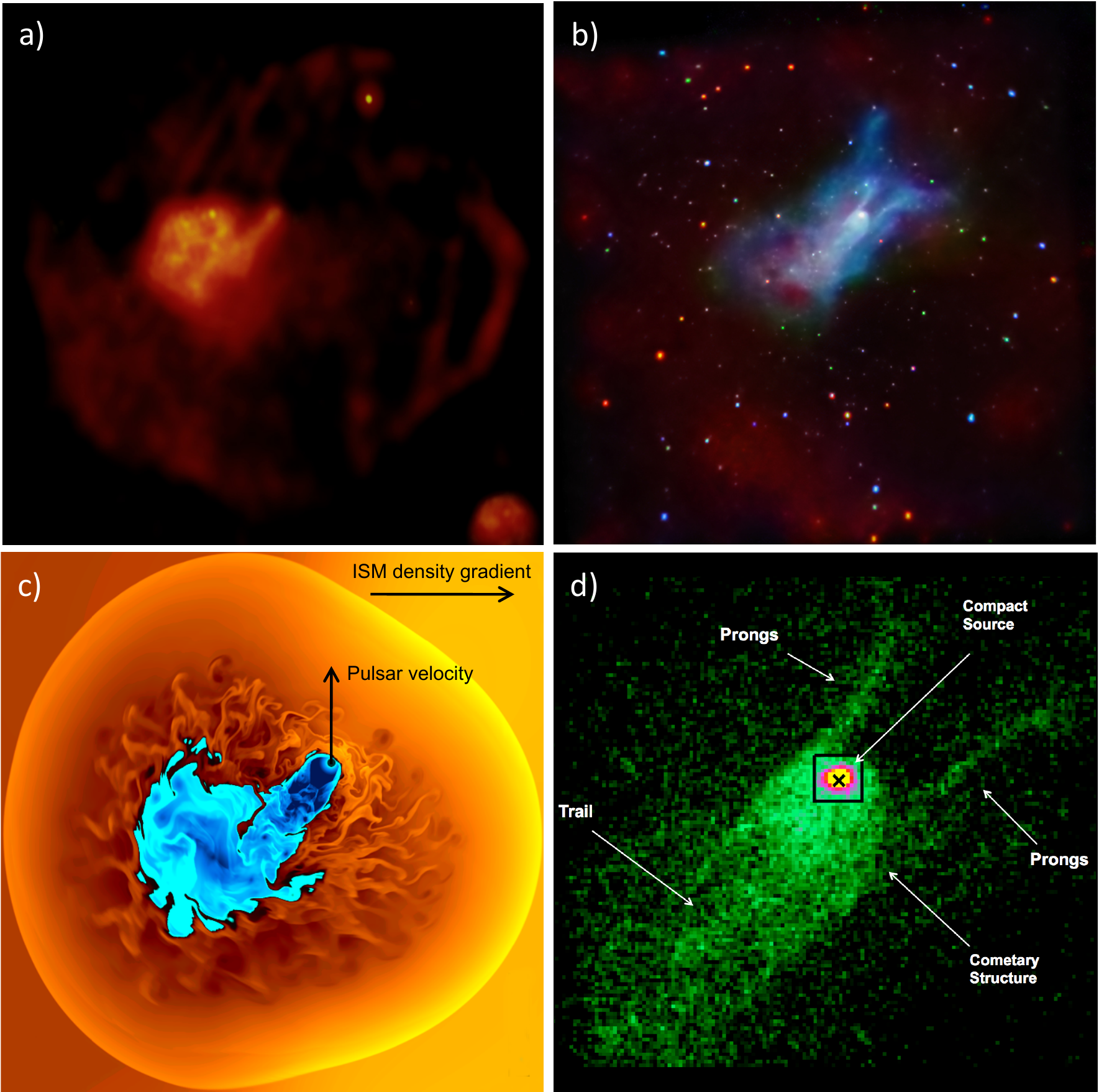}
\caption{
a) Radio emission from G327.1$-$1.1 (SIFA/MOST, CSIRO/ATNF/ATCA)
showing SNR shell surrounding central PWN. b) \chandra\ image 
showing faint X-ray SNR shell and PWN. c) Hydrodynamical simulation
of evolved composite SNR with properties similar to G327.1$-$1.1.
(See text for details.) d) Expanded \chandra\ view of central region
of G327.1$-$1.1. A compact nebula surrounding the neutron star is
embedded in a cometary structure with an extended tail, formed by
a combinition of northward pulsar motion and an interaction with
the SNR reverse shock approaching from the west. Prong-like
structures of unknown origin extend from several regions around the
nebula. [After Temim et al. 2015. All images have north at top and
west at the right.]
}
\label{fig:figure7}       
\end{figure}

An example of such a RS-interaction stage is presented in
Figure~\ref{fig:figure7} where we show the composite SNR G327.1$-$1.1
(Temim et al.  2015). Radio observations (a) show a complete SNR
shell surrounding an extended flat-spectrum PWN in the remnant
interior, accompanied by a finger-like structure extending to the
northwest. X-ray observations (b) show faint emission from the SNR
shell along with a central compact source located at the tip of the
radio finger, accompanied by a tail of emission extending back into
the radio PWN. The X-ray properties of the compact source are consistent
with emission from a pulsar (though, to date, pulsations have not
yet been detected) which, based on its position relative to the
geometric center of the SNR, appears to have a northward motion.
Spectra from the SNR shell indicate a density gradient in the
surrounding medium, increasing from east to west. Results from
hydrodynamical modeling of the evolution of such a system using
these measurements as constraints, along with an estimate for the
spin-down power of the pulsar based upon the observed X-ray emission
of its PWN (see Section 3.1) are shown in Figure~\ref{fig:figure7}c
where we show the density (compare with Figure~\ref{fig:figure3}).
The RS has approached rapidly from the west, sweeping past the
pulsar and disrupting the PWN. The result is a trail of emission
swept back into the relic PWN, in excellent agreement with the radio
morphology. The X-ray spectrum of the tail shows a distinct steepening
with distance from the pulsar, consistent with synchrotron cooling
of the electrons based on the estimated magnetic field and age of
the injected particles tracked in the hydro simulation. Detailed
investigation shows that the central source is actually resolved,
suggesting that the pulsar is surrounded by a compact nebula (panel
d). This is embedded in a cometary structure produced by a combination
of the northward motion of the pulsar and the interaction with the
RS propagating from the west. However, extended prong-like structures
are observed in X-rays, whose origin is currently not understood.

\subsection{Late-phase Evolution} \index{late-phase evolution}

As illustrated in Figure~\ref{fig:figure2}, as a PWN ages, the ratio
of the IC to synchrotron luminosity increases due to the declining
magnetic field in the nebula. As a result, in late phases of the
evolution, the $\gamma$-ray emission may dominate that observed in
the radio or X-ray bands. Indeed, PWNe dominate the population of
TeV $\gamma$-ray sources in the Galactic Plane (e.g., Carrigan et
al. 2013).  For many such TeV-detected PWNe, the inferred magnetic
field strengths are only $\sim 5 \ \mu$G (e.g., de Jager et al.
2008). In such a case, 1~TeV gamma-rays originate from electrons
with energies of $\sim 20$~TeV (assuming IC scattering of CMB
photons) while 1~keV synchrotron X-rays originate from electrons
with energies of $\sim 100$~TeV (see Eqns. \ref{eqn:E_syn},
\ref{eqn:E_IC}). The higher energy X-ray producing electrons fall
beyond the cooling break, while those producing the $\gamma$-rays
are predominantly uncooled. The result is a bright TeV nebula
accompanied by a fainter X-ray nebula.

Such results are seen clearly for HESS J1825$-$137, for which measurements
show that the TeV emission extends to much larger distances than 
the X-ray emission due to more rapid cooling of the X-ray emitting
particles. Indeed, for this PWN, the $\gamma$-ray size is observed
to decline with increasing energy, indicating that even some of the
$\gamma$-ray emitting electrons fall beyond the cooling break although,
as observed in younger PWNe in X-rays, the high energy emission extends
to larger radii than can be explained unless rapid diffusion of the
associated electrons is occurring (Van Etten \& Romani 2011). Deep surveys
with future VHE $\gamma$-ray telescopes are expected to reveal many
older systems for which emission in other wavebands is now faint.

\subsection{Escaping the SNR -- Bow Shock PWNe} \index{bow shock nebulae}
\label{sec:4.3}

\begin{figure}[t]
\sidecaption
\includegraphics[width=11.5cm]{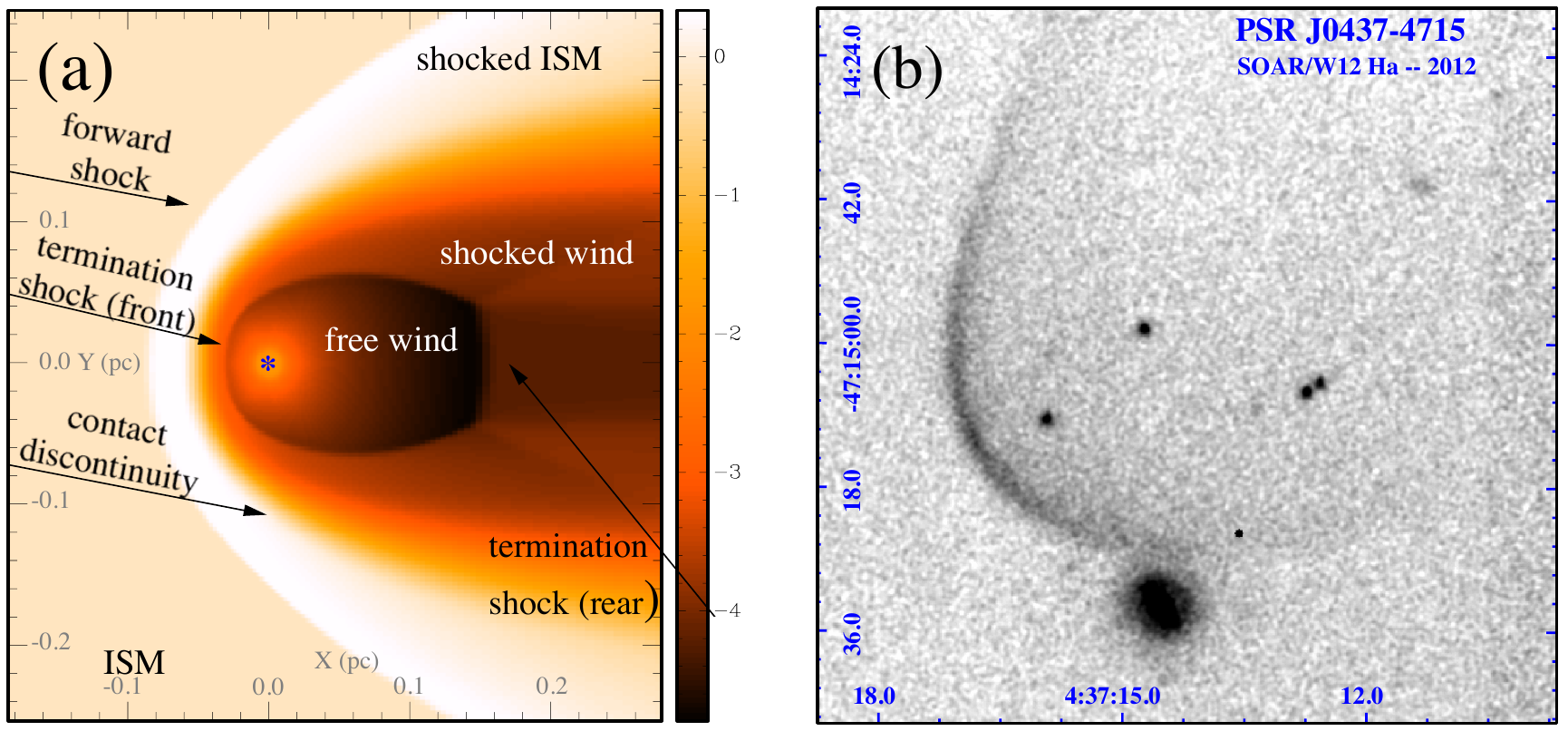}
\caption{
Left: Hydrodynamic simulation of a pulsar bow shock nebula (see
text). [From Gaensler \& Slane 2006. Reprinted by permission.]
Right: H$\alpha$ image of bow shock created by PSR~J0437$-$4715.
[Image courtesy of
R. Romani. See Brownsberger \& Romani 2014.]
}
\label{fig:figure8}       
\end{figure}

Late in the evolution of a PWN, the pulsar will exit its host SNR
and begin traveling through the ISM. Since the sound speed for the
cold, warm, and hot phases of the ISM is $v_s \sim 1,\ 10,\ \rm\
and\ 100 {\rm\ km\ s}^{-1}$,  the pulsar motion will be supersonic.
The relative motion of the ISM sweeps the pulsar wind back into a
bow shock structure. As illustrated in Figure~\ref{fig:figure8}
(left), the structure is still characterized by an FS, CD, and TS,
but the gas behind the FS is now shocked ISM material, and the CD
separates the shocked pulsar wind from the shocked ISM. Inside the
TS, the pulsar wind flows freely. The distance from the pulsar to
the TS depends on the angle $\theta$ relative to the pulsar motion
(as does that to the FS), and is approximately described by (Wilkin
1996)
\begin{equation}
R_w(\theta) = R_{w0} \frac{\sqrt{3(1-\theta \cot\theta)}}{\sin \theta}.
\end{equation}
Here $R_{w0}$ is the stand-off distance from the pulsar, in the 
direction of motion, where the wind pressure matches the ram
pressure of the inflowing ISM (in the pulsar frame):
\begin{equation}
R_{w0} = \sqrt{\dot{E}/(r \pi \omega c \rho_0 v_{\rm PSR}^2},
\end{equation}
where $v_{\rm PSR}$ is the pulsar velocity and $\rho_0$ is the
density of the unshocked ISM (cf. Equation 6).  Although this description 
was derived for a non-relativistic, unmagnetized radiative fluid,
while the pulsar wind is magnetized and relativistic, and the radiative
time for the ISM is long relative to the flow timescale in pulsar
bow shock nebulae, the overall geometric description provides
an adequate representation (Bucciantini \& Bandiera 2001).

The non-radiative shock formed in the ISM interaction results in
the emission of optical Balmer lines, dominated by H$\alpha$,
providing a distinct signature from which properties of the pulsar
motion and wind luminosity can be inferred. An exceptional example
is the bow shock nebula associated with PSR~J0437$-$4715
(Figure~\ref{fig:figure8}, right), a nearby ms pulsar in a binary
system, for which timing measurements have established $M_{NS} \sim
1.8 M_\odot$ (Verbiest et al. 2008). Parallax measurements establish
a distance of 0.16~kpc, and proper motion measurements of the pulsar
(and nebula) provide $v_\perp = 107 {\rm\ km\ s}^{-1}$.  With the
measured spin-down power $\dot{E} = 5.5 \times 10^{33} {\rm\ erg\
s}^{-1}$, modeling of the bow shock structure provides a direct
limit on the NS moment of inertia that indicates a relatively stiff
equation of state (Brownsberger \& Romani 2014).

Radio and X-ray measurements of bow shock nebulae probe the shocked
pulsar wind. Observations of PSR~J1747$-$2958 and its associated
nebula G359.23$-$0.82 reveal a long radio tail and an X-ray morphology
that reveals both a highly magnetized tail from wind shocked from
the forward direction, and a weakly magnetized tail from wind flowing
in the direction opposite that of the pulsar motion (Gaensler et al.
2004). High resolution measurements of the emission near
several pulsars have also provided evidence for asymmetric pulsar winds
imprinting additional structure on the bow shock structure (e.g.,
Romani et al. 2010).

\section{Summary}
\label{sec:5}

The structure of a PWN is determined by both the properties of the
host pulsar and the environment into which the nebula expands.
Observations across the electromagnetic spectrum allow us to constrain
the nature of the pulsar wind, including both its magnetization and
geometry, and the global properties of the PWN allow us to constrain
the evolutionary history as it evolves through the ejecta of the
supernova remnant in which it was born. Spectroscopic observations
yield information on the mass and composition of shocked ejecta
into which the nebula expands, and on the expansion velocity.
Measurements of the broadband spectrum provide determinations of
the nebular magnetic field and the maximum energy of the particles
injected into the PWN.  These observations continue to inform
theoretical models of relativistic shocks which, in turn, have broad
importance across the realm of high-energy astrophysics.

At late phases, interactions between the PWN and the SNR RS produce
a complex combination of the relic nebula and freshly-injected
particles. Hydrodynamical simulations of the entire composite SNR
system can reveal information on the long-term evolution, which
depends on the details of the pulsar motion, its wind properties,
the ejecta mass and explosion energy of the SNR, and the properties
of the surrounding medium. Such systems may eventually fade into
obscurity, with $\gamma$-ray emission from the relic electrons
providing an important signature before the pulsars exit their SNRs
and traverse the ISM at supersonic speeds, producing elongated bow
shock nebulae whose structure continue to provide a glimpse of the
relativistic outflows from the aging pulsars.

\bigskip
\noindent
{\large{\bf Acknowledgements}}

\noindent
The author would like to thank the many colleagues with whom he
has collaborated on studies that have been briefly summarized in
this Handbook contribution. Partial support for this effort was
provided by NASA Contract NAS8-03060.

\bigskip

\noindent
{\large{\bf Cross-References}}

\noindent
$\bullet$ Supernova of 1054 and its remnant, the Crab Nebula

\noindent
$\bullet$ The Historical Supernova of AD1181 and its remnant, 3C58

\noindent
$\bullet$ Supernovae from super AGB Stars (8-12 $M_\odot$)

\noindent
$\bullet$ Explosion Physics of Core - Collapse Supernovae

\noindent
$\bullet$ Radio Neutron Stars

\noindent
$\bullet$ Distribution of the spin periods of neutron stars

\noindent
$\bullet$ Dynamical Evolution and Radiative Processes of Supernova Remnants

\noindent
$\bullet$ X-ray Emission Properties of supernova remnants

\noindent 
$\bullet$ Infrared Emission from Supernova Remnants:
Formation and Destruction of Dust

%
%
\bibliographystyle{spbasic}
\bibliography{ref}

\end{document}